\documentclass[11pt]{article}
\usepackage{graphicx,color}
\begin{document}

\title{
Study on Decays of $D_{sJ}^{*}(2317)$ and $D_{sJ}(2460)$ in terms
of the CQM Model} \maketitle

\centerline{Xiang Liu,  Yan-Ming Yu, Shu-Min Zhao and Xue-Qian Li}

\vspace{0.8cm}

\centerline{Department of Physics, Nankai University, Tianjin
300071, China}

\vspace{0.8cm}

\begin{abstract}
Based on the assumption that $D_{sJ}^{*}(2317)$ and $D_{sJ}(2460)$
are the $(0^+ ,1^+)$ chiral partners of  $D_{s}$ and $D^* _s$, we
evaluate the strong pionic  and radiative decays of
$D_{sJ}^{*}(2317)$ and $D_{sJ}(2460)$ in the Constituent Quark
Meson (CQM) model.  Our numerical results of the relative ratios
of the decay widths are reasonably consistent with data.

\end{abstract}


\section{Introduction}

The new discoveries of exotic particles $D_{sJ}^{*}(2317)$ and
$D_{sJ}(2460)$, which possess spin-parity structures of $0^+$,
$1^+$ respectively \cite{exp,CLEO,exp1}, attract great interests
of both theorists and experimentalists of high energy physics.
Some authors \cite{Bardeen} suppose that $D_{sJ}^{*}(2317)$ and
$D_{sJ}(2460)$ are  $(0^+,1^+)$ chiral partners of $D_s$ and
$D^*_s$ i.e. p-wave excited states of $D_s$ and $D^*_s$
\cite{Chao}. Narison used the QCD spectral sum rules to calculate
the masses of $D_{sJ}^{*}(2317)$ and $D_{sJ}(2460)$ by assuming
that they are quark-antiquark states and obtained results which
are consistent with data within a wide error range \cite{Narison}.
Beveren and Rupp also studied the mass spectra \cite{Rupp} and
claimed that their results support the $c\bar s$ structures for
$D_{sJ}^{*}(2317)$ and $D_{sJ}(2460)$. Meanwhile, some other
authors suggest that $D_{sJ}^{*}(2317)$ and $D_{sJ}(2460)$ can
possibly be four-quark states \cite{four,decay-5,mole,atom}. Thus,
one needs to try various ways to understand the structures of
$D_{sJ}^{*}(2317)$ and $D_{sJ}(2460)$. In general, one can take a
reasonable theoretical approach to evaluate related physical
quantities and then compare the results with data to extract
useful information. One can determine the structures of
$D_{sJ}^{*}(2317)$ and $D_{sJ}(2460)$ by studying the production
rates of the exotic particles, and our recent work \cite{liu} is
just about the production of $D_{sJ}^{*}(2317)$ in the decays of
$\psi(4415)$.

Recently, several groups have calculated the  strong and radiative
decay rates of $D_{sJ}^{*}(2317)$ and $D_{sJ}(2460)$ in different
theoretical approaches:  the Light Cone QCD Sum Rules, Constituent
Quark model, Vector Meson Dominant (VMD) ansatz, etc.
\cite{Nielsen,zhu,decay-1,decay-2,decay-3,decay-4,radi-LCQSR}. For
a clear comparison, the results by different groups are listed in
Tables 2 and 3. The authors of Ref. \cite{decay-5,decay-6} also
calculated the rates based on the assumption that
$D_{sJ}^{*}(2317)$ and $D_{sJ}(2460)$ are in non-$c\bar s$
structures. Their predictions on the $D_{sJ}^{*}(2317)$ and
$D_{sJ}(2460)$ decay rates are obviously larger than that obtained
by assuming the two-quark structure by orders. Thus studies on the
strong and radiative decays with other plausible models would be
helpful. It can not only deepen our understanding about the
characters of these particles, but also test the reliability of
models which are applied to calculate the decays. Because
$D_{sJ}(2632)$ was only observed by the SELEX
collaboration\cite{selex}, but not by Babar \cite{0408087}, Belle
\cite{2632-belle} and FOCUS\cite{Focus}, its existence is still in
dispute, so here we do not refer decays of $D_{sJ}(2632)$.

In this work, we study the strong and radiative decays of
$D_{sJ}^{*}(2317)$ and $D_{sJ}(2460)$ in the framework of
Constituent-Quark-Meson (CQM) model. CQM model was proposed by
Polosa et al. \cite {CQM} and has been  well developed later based
on the works of Ebert et al. \cite{feldmann} (See the Ref.
\cite{CQM} for a review). The model is based on an effective
Lagrangian which incorporates the flavor-spin symmetry for heavy
quarks with the chiral symmetry for light quarks. Employing the
CQM model to study phenomenology of heavy meson physics,
reasonable results have been achieved
\cite{application,parameter}. Therefore, we can believe that the
model is applicable to our processes and expect to get relatively
reliable conclusion.

The constraint from the phase space of the final states forbids
the processes $D_{sJ}^{*}(2317)\rightarrow D_{s}\eta (\eta')$ and
$D_{sJ}(2460)\rightarrow D_{s}^{*}\eta (\eta')$, so that the only
allowed strong decay modes are $D_{sJ}^{*}(2317)\rightarrow
D_{s}\pi^{0}$ and $D_{sJ}(2460)\rightarrow D_{s}^{*}\pi^0$. In
principle, $D_{sJ}(2460)\rightarrow D_{sJ}^{*}(2317)+\pi^0$ which
is a $1^+\rightarrow 0^++0^-$ process, is allowed by the phase
space. However, it is a p-wave reaction, and the total rate is
proportional to $|{\bf p}|^2$, where ${\bf p}$ is the
three-momentum of emitted pion in the center-of-mass frame of
$D_{sJ}(2460)$. In this case $|{\bf p}|$ is very small (about
$\sim 28$ MeV), so that this process can only contribute to the
total width a negligible fraction, in practice.

The aforementioned strong decay modes obviously violate isospin
conservation. Therefore the decay widths of
$D_{sJ}^{*}(2317)\rightarrow D_{s}\pi^{0}$ and
$D_{sJ}(2460)\rightarrow D_{s}^{*}\pi^0$ must be highly
suppressed. Moreover, direct emission of a pion is OZI suppressed
\cite{OZI}.

Cho et al. suggested a mixing mechanism of  $\eta-\pi^0$ where the
isospin violation originates from the mass splitting  of $u$ and
$d$ quarks \cite{wise}. In that scenario, $D_{sJ}^{*}(2317)$ and
$D_{sJ}(2460)$ firstly transit into $ D_{s}\eta$ and
$D_{s}^{*}\eta$, and then $\eta$ transits into $\pi^0$ by the
mixing. In the intermediate process, $\eta$ obviously is
off-shell. The mixing depends on the mass difference of $\eta$ and
$\pi$, and the effects due to the mixing between $\eta'$ and $\pi$
can be ignored.

Another sizable mode is the radiative decay. Even though, by
general consideration, the electromagnetic reaction should be much
less important than the strong decay, it does not suffer the
suppression of isospin violation, therefore one may expect that it
has a comparable size to the strong processes described above. The
relevant decay modes are $D_{sJ}^{*}(2317)\rightarrow
D_{s}^{*}+\gamma$, $D_{sJ}(2460)\rightarrow D_{s}+\gamma$ and
$D_{sJ}(2460)\rightarrow D_{s}^* +\gamma$ and
$D_{sJ}(2460)\rightarrow D_{sJ}^{*}(2317)+\gamma$.

Currently the Babar  and the Belle collaborations have completed
precise measurements on the ratio of
$\Gamma(D_{sJ}(2460)\rightarrow D_{s}\gamma)$ to
$\Gamma(D_{sJ}(2460)\rightarrow D_{s}^* \pi^0)$
\cite{exp1,belle,babar-1}.  Now Babar  and Belle collaborations
begin to further measure other decays of $D_{sJ}^{*}(2317)$ and
$D_{sJ}(2460)$. We are expecting new results of Babar and Belle
which can be applied to decisively determine the structures of
$D_{sJ}^{*}(2317)$ and $D_{sJ}(2460)$.

This paper is organized as follows, after the introduction, in
Sect. 2, we formulate the strong and radiative decays of
$D_{sJ}^{*}(2317)$ and $D_{sJ}(2460)$. In Sect.3, we present our
numerical results along with all the input parameters. Finally,
Sect. 4 is devoted to discussion and conclusion. Some detailed
expressions are collected in the Appendix.

\section{Formulation}

First, for readers' convenience, we present a brief introduction
of the Constituent-Quark-Meson (CQM) model \cite{CQM}. The model
is relativistic and based on an effective Lagrangian which
combines the heavy quark effective theory (HQET) and the chiral
symmetry for light quarks,
\begin{eqnarray}
L_{CQM}&=&\bar{\chi}[\gamma\cdot (i\partial+\mathcal{V})]\chi
+\bar{\chi}\gamma\cdot\mathcal{A}\gamma_{5}\chi-m_{q}\bar{\chi}\chi+\frac{f_{\pi}^2}{8}
\mathrm{Tr}[\partial^{\mu}\Sigma\partial_{\mu}\Sigma^{+}]+\bar{h}_{v}(iv\cdot
\partial)h_{v}\nonumber\\&&-[\bar{\chi}(\bar{H}+\bar{S}+i\bar{T}^{\mu}\frac{\partial_{\mu}}
{\Lambda})h_{v}+h.c.]+\frac{1}{2G_{3}}\mathrm{Tr}[(\bar{H}+\bar{S})(H-S)]+\frac{1}{2G_{4}}\mathrm{Tr}
[\bar{T}^{\mu}T_{\mu}],
\end{eqnarray}
where the fifth term is the kinetic term of heavy quarks with
$v\!\!\!\slash h_v=h_v$; $H$ is the super-field corresponding to
doublet $(0^{-},1^-)$ of negative parity  and has an explicit
matrix representation:
$$H=\frac{1+v\!\!\!\slash}{2}(P_{\mu}^{*}\gamma^{\mu}-P\gamma_{5});$$
where $P$ and $P^{*\mu}$ are the annihilation operators of
pseudoscalar and vector mesons which are normalized as
$$\langle
0|P|M(0^-)\rangle=\sqrt{M_{H}},\;\;\;\;\;{\rm and}\;\;\;\;\;\;
\langle 0|P^{*\mu}|M(1^-)\rangle=\sqrt{M_{H}}\epsilon^{\mu}.$$

$S$ is the super-fields related to $(0^{+},1^{+})$,
$$S=\frac{1+v\!\!\!\slash}{2}[P^{*'}_{1\mu}\gamma^{\mu}\gamma_{5}-P_{0}].$$
$\chi=\xi q$ ($q=u,d,s$) is the light quark field and
$\xi=e^{\frac{i\mathcal{M}}{f_{\pi}}}$, and $M$ is the octet
pseudoscalar matrix. We also have
$$\mathcal{V}^{\mu}=\frac{1}{2}(\xi^{\dag}\partial^{\mu}\xi+\xi\partial^{\mu}\xi^{\dag}),$$
and
$$\mathcal{A}^{\mu}=\frac{-i}{2}(\xi^{\dag}\partial^{\mu}\xi-\xi\partial^{\mu}\xi^{\dag}).$$

Because the spin-parity of $D_{sJ}^{*}(2317)$ and $D_{sJ}(2460)$
are $0^{+}$ and $1^+$, thus $D_{sJ}^{*}(2317)$ and $D_{sJ}(2460)$
can be embedded into the S-type doublet $(0^{+}, 1^{+})$
\cite{parameter}, whereas $D_{s}$ and $D_{s}^{*}$ belong to the H
type doublet $(0^{-}, 1^{-})$. Then we can calculate the strong
and radiative decay rates of $D_{sJ}^{*}(2317)$ and $D_{sJ}(2460)$
in the CQM model.

\subsection{The transition amplitude of $D_{sJ}^{*}(2317)\rightarrow D_{s}+\pi^0$
and $D_{sJ}(2460)\rightarrow D^*_{s}+\pi^0$ strong decays in the
CQM model.}

\begin{figure}[htb]
\begin{center}
\scalebox{0.8}{\includegraphics{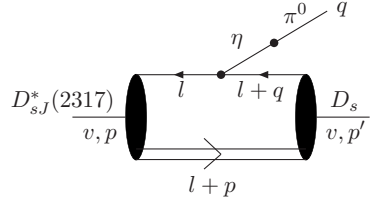}}
\end{center}
\caption{The Feynman diagrams which depict the decays of
$D_{sJ}^{*}(2317)\rightarrow D_{s}+\pi^0$ or
$D_{sJ}(2460)\rightarrow D_{s}^{*}+\pi^0$. The double-line denotes
the heavy quark ($c-$quark) propagator.} \label{diagram}
\end{figure}

As discussed above indirect $D_{sJ}^{*}(2317)\rightarrow
D_{s}+\pi^0$ and $D_{sJ}(2460)\rightarrow D_{s}^*+\pi^0$  occur
via two steps. In Fig. \ref{diagram}, we show the Feynman diagrams
which depict the strong processes $D_{sJ}^{*}(2317)\rightarrow
D_{s}+\eta\rightarrow D_{s}+\pi^0$ and $D_{sJ}(2460)\rightarrow
D_{s}^{*}+\eta\rightarrow D_{s}^{*}+\pi^0$. According to chiral
symmetry, $\eta$ only couples with light quark, thus it can only
be emitted from the light-quark leg in Fig.\ref{diagram}.

The matrix elements of $D_{sJ}^{*}(2317)\rightarrow D_{s}+\pi^0$
and $D_{sJ}(2460)\rightarrow D_{s}^{*}+\pi^0$ are written as
\begin{eqnarray}
&& \mathcal{M}[D_{sJ}^{*}(2317)\rightarrow D_{s}+\pi^0]=\langle
\pi^0 |\mathcal{L}_{mixing}|\eta\rangle \frac{i}{m_{\eta}^2
-m_{\pi}^2}\langle\eta D_{s}|\mathcal{L}_{CQM}|
D_{sJ}^{*}(2317)\rangle,\\
&& \mathcal{M}[D_{sJ}(2460))\rightarrow D^{*}_{s}+\pi^0]=\langle
\pi^0 |\mathcal{L}_{mixing}|\eta\rangle \frac{i}{m_{\eta}^2
-m_{\pi}^2}\langle\eta D_{s}^{*}|\mathcal{L}_{CQM}|
D_{sJ}(2460)\rangle.\label{la}
\end{eqnarray}

The mixing mechanism is described by the Lagrangian\footnote{Here
we ignore the mixing of $\eta$ and $\eta'$, because the mixing
angle $\theta\sim -11^{\circ}$ is small and does not much affect
our results. Therefore we simply assume that $\eta$ is $\eta_8$
and the contribution of $\eta'$, as discussed in the text, is
neglected in our calculations. }
$$\mathcal{L}_{mixing}=\frac{m_{\pi}^2 }{2\sqrt{3}}\frac{
\tilde{m}_{u}-\tilde{m}_{d}}{\tilde{m}_{u}+\tilde{m}_{d}}\pi^{0}\eta$$
which originates from the mass term of the low energy Lagrangian
for the pseudoscalar octet \cite{wise}
\begin{eqnarray}
\mathcal{L}_{mass}=\frac{m^2_{\pi}f_{\pi}^2}{4(\tilde{m}_{u}+\tilde{m}_{d})}\mathrm{Tr}[\xi
m_{q}\xi+\xi^{\dag}m_{q}\xi^{\dag}],
\end{eqnarray}
where $m_{q}$ is the light quark mass matrix. The matrix elements
of $\langle\eta D_{s}|\mathcal{H}_{CQM}| D_{sJ}^{*}(2317)\rangle$
and $\langle\eta D_{s}^{*}|\mathcal{H}_{CQM}| D_{sJ}(2460)\rangle$
will be calculated in the CQM model.

$\tilde{m}_{i}(i=u,d,s)$ are the current quark masses.  It is
noted that in the CQM model calculations,  the quark masses
($m_q,\; m_c$) which we denote as $m_q$, are constituent quark
masses \cite{CQM}, whereas, for the mixing, the concerned masses
which we denote as $\tilde m_q$ are the current quark
masses\cite{wise}.

According to the CQM model  \cite{CQM}, couplings of
$D_{sJ}^{*}(2317)$ and $D_{sJ}(2460)$  with light and heavy quarks
are expressed as
$$\frac{1+v\!\!\!\slash}{2}\sqrt{Z_{S}M_{1}},$$ and
$$\frac{1+v\!\!\!\slash}{2}\sqrt{Z_{S}M_{2}}\epsilon\!\!\!\slash_{1}
\gamma_{5},$$ the couplings of $D_{s}$ and $D_{s}^*$ to light and
heavy quarks are
$$\frac{1+v\!\!\!\slash}{2}\sqrt{Z_{S}M_{D_{s}}}\gamma_{5}$$ and
$$\frac{1+v\!\!\!\slash}{2}\sqrt{Z_{S}M_{D_{s}^{*}
}}\epsilon\!\!\!\slash_{2},$$ where $\epsilon_{1}$ and
$\epsilon_{2}$ denote the polarization vectors of $D_{sJ}(2460)$
and $D_{s}^{*}$ respectively. $M_{1}$ and $M_2$ are respectively
the masses of $D_{sJ}^{*}(2317)$ and $D_{sJ}(2460)$.  The concrete
expressions of $Z_{H}$ and $Z_{S}$ are given in Ref. \cite{CQM} as
\begin{eqnarray}
Z_{H}^{-1}&=&(\Delta_{H}+m_{s})\frac{\partial
I_{3}(\Delta_{H})}{\partial\Delta_{H}}+ I_{3}(\Delta_{S}),\label{ZH}\\
Z_{S}^{-1}&=&(\Delta_{S}+m_{s})\frac{\partial
I_{3}(\Delta_{S})}{\partial\Delta_{S}}+ I_{3}(\Delta_{S}),\label{ZS}\\
I_{3}(a)&=&\frac{iN_c}{16\pi^4}\int^{1/\mu^2}_{1/\Lambda^2}\frac{dy}{y^{3/2}}
\exp[-y(m_{s}^{2}-a^2)](1+\mathrm{erf}(a\sqrt{y})),
\end{eqnarray}
where erf is the error function.

Now, we can write out the transition matrix elements as
\begin{eqnarray}
&&\langle\eta D_{s}|\mathcal{H}_{CQM}|
D_{sJ}^{*}(2317)\rangle\nonumber\\&=& (-1)i^6
\sqrt{Z_{S}M_{1}Z_{H}M_{D_{s}}}\sqrt{\frac{2}{3}}\frac{N_{c}}{2f_{\pi}}\int^{reg}\frac{d^4
l}{(2\pi)^4}\frac{\mathrm{Tr}[(l\!\!\!\slash+m_{s})q^{\mu}\gamma_{\mu}\gamma_{5}
(l\!\!\!\slash+q\!\!\!\slash+m_{s})\gamma_{5}
(1+v\!\!\!\slash)]}{(l^2-m_{s}^2)[(l+q)^2-m_{s}^2](v\cdot
l+\Delta_{S})},\nonumber\\
\end{eqnarray}
and
\begin{eqnarray}
&&\langle\eta D_{s}^{*}|\mathcal{H}_{CQM}|
D_{sJ}(2460)\rangle\nonumber\\&=& (-1)i^6
\sqrt{Z_{S}M_{2}Z_{H}M_{D_{s}^*}}\sqrt{\frac{2}{3}}\frac{N_{c}}{2f_{\pi}}
\int^{reg}\frac{d^4
l}{(2\pi)^4}\frac{\mathrm{Tr}[(l\!\!\!\slash+m_{s})q^{\mu}\gamma_{\mu}\gamma_{5}
(l\!\!\!\slash+q\!\!\!\slash+m_{s})\epsilon\!\!\!\slash_{2}
(1+v\!\!\!\slash)\gamma_{5}
\epsilon\!\!\!\slash_{1}]}{(l^2-m_{s}^2)[(l+q)^2-m_{s}^2](v\cdot
l+\Delta_{S})},\nonumber\\
\end{eqnarray}
where $N_{c}=3$.

Omitting  technical details in the text for saving space, we
finally obtain
\begin{eqnarray}
&&\langle\eta D_{s}|\mathcal{H}_{CQM}|
D_{sJ}^{*}(2317)\rangle=\sqrt{Z_{S}M_{1}Z_{H}M_{D_{s}}}
\Big(-\sqrt{\frac{2}{3}}\Big)\frac{\mathcal{A}}{2f_{\pi}},\\
&&\langle\eta D_{s}^{*}|\mathcal{H}_{CQM}| D_{sJ}(2460)\rangle=
\sqrt{Z_{S}M_{2}Z_{H}M_{D_{s}^*}}\Big(-\sqrt{\frac{2}{3}}\Big)\frac{(\epsilon_{1}
\cdot\epsilon_{2})\mathcal{A}} {2f_{\pi}},
\end{eqnarray}
with
\begin{eqnarray}
\mathcal{A}=4(m_{s}^2 m_{\eta} \omega -m_{\eta}^2
m_{s})\mathcal{O}-m_{\eta}^2 (\mathcal{O}_{1}+\omega
\mathcal{O}_{2})+m_{\eta}\omega(2
\mathcal{O}_{3}-\mathcal{O}_{4}-\mathcal{O}_{5}-2\mathcal{O}_{6}),
\end{eqnarray}
where $\omega=v\cdot v'=\frac{\Delta_{S}-\Delta_{H}}{2m_{\eta}}$
and  the concrete expressions of $\mathcal{O}$ and
$\mathcal{O}_{i}$ are listed in the appendix.

The decay widths of $D_{sJ}^{*}(2317)\rightarrow D_{s}+\pi^0$ and
$D_{sJ}(2460)\rightarrow D^*_{s}+\pi^0$ read as
\begin{eqnarray}
&& \Gamma[D_{sJ}^{*}(2317)\rightarrow
D_{s}+\pi^0]\nonumber\\&&=\frac{Z_{S}Z_{H}M_{D_{s}}|\textbf{p}'|}{1024\pi
M_{1}f_{\pi}^{2}}\bigg(\frac{\tilde{m}_u -\tilde{m}_d
}{\tilde{m}_{s}-(\tilde{m}_{u}+\tilde{m}_{d})/2}\bigg)^2
|\mathcal{A}|^2,\\
&& \Gamma[D_{sJ}(2460)\rightarrow
D_{s}^{*}+\pi^0]\nonumber\\&&=\frac{Z_{S}Z_{H}M_{D_{s}}|\textbf{p}'|}{3072\pi
M_{2}f_{\pi}^{2}}\bigg(\frac{\tilde{m}_u -\tilde{m}_d
}{\tilde{m}_{s}-(\tilde{m}_{u}+\tilde{m}_{d})/2}\bigg)^2
\bigg[2+\frac{(M_{2}^{2}+M_{D_{s}^{*}}^{2})^2}{M_{2}^{2}M_{D_{s}^{*}}^{2}}\bigg]
|\mathcal{A}|^2,
\end{eqnarray}
where  the relations $m_{\pi}^2 =2\tilde{m}B_{0}$ and $m_{\eta}^2
=\frac{2}{3}(\tilde{m}+2\tilde{m}_{s})B_{0}$ with
$\tilde{m}=(\tilde{m}_{u}+\tilde{m}_{d})/2$ are employed to derive
isospin suppression factor $(\tilde{m}_u -\tilde{m}_d)/[
\tilde{m}_{s}-(\tilde{m}_{u}+\tilde{m}_{d})/2]$ \cite{coe}. These
relations are valid at the leading order of the chiral theory, but
the principle does not really apply for estimating the mass of
$\eta'$ (or $\eta_0$) due to an extra contribution from the axial
anomaly \cite{faessler}.

\begin{figure}[htb]
\begin{center}
\begin{tabular}{cc}
\scalebox{0.8}{\includegraphics{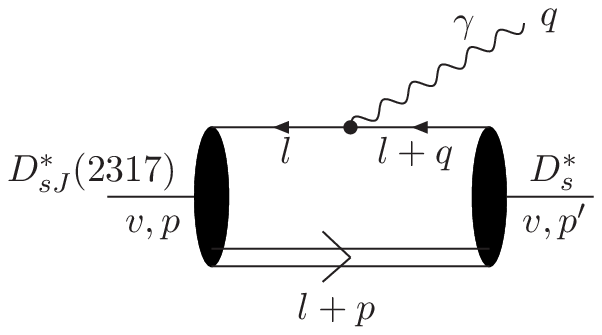}}&\scalebox{0.8}{\includegraphics{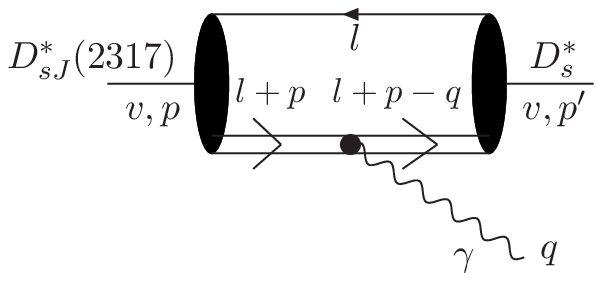}}\\
(a)&(b)\\
\end{tabular}
\end{center}
\caption{The Feynman diagrams depicts the radiative decays of
$D_{sJ}^{*}(2317)$ and $D_{sJ}(2460)$.} \label{diagram-1}
\end{figure}

\subsection{The transition amplitude of $D_{sJ}^{*}(2317)$ and $D_{sJ}(2460)$
radiative decays in the CQM model.}

the Feynman diagrams of $D_{sJ}^{*}(2317)\rightarrow
D_{s}^{*}+\gamma$, $D_{sJ}(2460)\rightarrow D_{s}+\gamma$,
$D_{sJ}(2460)\rightarrow D_{s}^* +\gamma$ and
$D_{sJ}(2460)\rightarrow D_{sJ}^{*}(2317)+\gamma$ are presented in
Fig. \ref{diagram-1}.
We will evaluate these radiative decays in the CQM model.
\vspace{0.5cm}

(a) $D_{sJ}^{*}(2317)\rightarrow D_{s}^{*}+\gamma$.

In contrast with the case in Fig.\ref{diagram},  the heavy quark
also couples to $\gamma$. Thus the transition matrix element of
$D_{sJ}^{*}(2317)\rightarrow D_{s}^{*}+\gamma$ is written as
\begin{eqnarray}
&&\mathcal{M}[D_{sJ}^{*}(2317)\rightarrow D_{s}^{*}+
\gamma]\nonumber\\&&= (-1)i^6
\sqrt{Z_{S}M_{1}Z_{H}M_{D^{*}_{s}}}N_{c}\bigg\{\bigg(\frac{-e}{3}\bigg)\int^{reg}\frac{d^4
l}{(2\pi)^4}\frac{\mathrm{Tr}[(l\!\!\!\slash+m_{s})\epsilon\!\!\!\slash_{\gamma}
(l\!\!\!\slash+q\!\!\!\slash+m_{s})\epsilon\!\!\!\slash_{2}
\frac{(1+v\!\!\!\slash)}{2}]}{(l^2-m_{s}^2)[(l+q)^2-m_{s}^2](v\cdot
l+\Delta_{S})}\nonumber\\&&+\frac{2e}{3}\int^{reg}\frac{d^4
l}{(2\pi)^4}\frac{\mathrm{Tr}[(l\!\!\!\slash+m_{s})\epsilon\!\!\!\slash_{2}\frac{(1+v\!\!\!\slash)}{2}
\epsilon\!\!\!\slash_{\gamma}\frac{(1+v\!\!\!\slash)}{2}]}{(l^2-m_{s}^2)(v\cdot
l+\Delta_{H})(v\cdot
l+\Delta_{S})}\bigg\}\nonumber\\
&&=\sqrt{Z_{S}M_{1}Z_{H}M_{D_{s}^{*}}}\Big(\frac{2e}{3}\Big)\Big\{
(\epsilon_{\gamma}\cdot \epsilon_{2})\Big[m_{s}^2 \mathcal{R}
-\mathcal{R}_{3}-2\mathcal{R}_{6}-\frac{(p'\cdot
q)(m_{s}\mathcal{R}+\mathcal{R}_{1}+2
\mathcal{R}_{5})}{M_{D_{s}^{*}}}\Big]\nonumber\\&&
+(\epsilon_{\gamma}\cdot p')(\epsilon_{2}\cdot
q)\Big[\frac{m_{s}\mathcal{R}+\mathcal{R}_{1}+2
\mathcal{R}_{5}}{M_{D_{s}^{*}}}\Big] \Big\}.
\end{eqnarray}

\vspace{0.5cm}
 (b) $D_{sJ}(2460)\rightarrow D_{s}+\gamma$.

The transition matrix element of $D_{sJ}(2460)\rightarrow
D_{s}+\gamma$ is

\begin{eqnarray}
&&\mathcal{M}[D_{sJ}(2460)\rightarrow D_{s}+\gamma ]\nonumber\\&&=
(-1)i^6
\sqrt{Z_{S}M_{2}Z_{H}M_{D_{s}}}N_{c}\bigg\{\bigg(\frac{-e}{3}\bigg)\int^{reg}\frac{d^4
l}{(2\pi)^4}\frac{\mathrm{Tr}[(l\!\!\!\slash+m_{s})\epsilon\!\!\!\slash_{\gamma}
(l\!\!\!\slash+q\!\!\!\slash+m_{s}+i\epsilon)\gamma_5{}
\frac{(1+v\!\!\!\slash)}{2}\gamma_{5}\epsilon\!\!\!\slash_{1}]}{(l^2-m_{s}^2)[(l+q)^2-m_{s}^2](v\cdot
l+\Delta_{S})}\nonumber\\&&+\frac{2e}{3}\int^{reg}\frac{d^4
l}{(2\pi)^4}\frac{\mathrm{Tr}[(l\!\!\!\slash+m_{s})\gamma_{5}\frac{(1+v\!\!\!\slash)}{2}
\epsilon\!\!\!\slash_{\gamma}\frac{(1+v\!\!\!\slash)}{2}\gamma_{5}\epsilon\!\!\!\slash_{1}]}{(l^2-m_{s}^2)(v\cdot
l+\Delta_{H})(v\cdot
l+\Delta_{S})}\bigg\}\nonumber\\
&&=\sqrt{Z_{S}M_{1}Z_{H}M_{D_{s}}}\Big(\frac{2e}{3}\Big)\Big\{
(\epsilon_{\gamma}\cdot \epsilon_{1})\Big[m_{s}^2 \mathcal{R}
-\mathcal{R}_{3}-2\mathcal{R}_{6}-\frac{(p'\cdot
q)(m_{s}\mathcal{R}+\mathcal{R}_{1}+2
\mathcal{R}_{5})}{M_{D_{s}}}\Big]\nonumber\\&&
+(\epsilon_{\gamma}\cdot p')(\epsilon_{1}\cdot
q)\Big[\frac{m_{s}\mathcal{R}+\mathcal{R}_{1}+2
\mathcal{R}_{5}}{M_{D_{s}}}\Big] \Big\}.
\end{eqnarray}

\vspace{0.5cm}

(c) $D_{sJ}(2460)\rightarrow D_{s}^{*}+\gamma$.

The transition matrix element of $D_{sJ}(2460)\rightarrow
D_{s}^{*}+\gamma$  is

\begin{eqnarray}
&&\mathcal{M}[D_{sJ}(2460)\rightarrow D_{s}^*+\gamma
]\nonumber\\&&= (-1)i^6
\sqrt{Z_{S}M_{2}Z_{H}M_{D_{s}^{*}}}{N_{c}}\bigg\{\bigg(\frac{-e}{3}\bigg)\int^{reg}\frac{d^4
l}{(2\pi)^4}\frac{\mathrm{Tr}[(l\!\!\!\slash+m_{s})\epsilon\!\!\!\slash_{\gamma}
(l\!\!\!\slash+q\!\!\!\slash+m_{s})\epsilon\!\!\!\slash_{2}
\frac{(1+v\!\!\!\slash)}{2}\gamma_{5}\epsilon\!\!\!\slash_{1}]}{(l^2-m_{s}^2)[(l+q)^2-m_{s}^2](v\cdot
l+\Delta_{S})}\nonumber\\&&+\frac{2e}{3}\int^{reg}\frac{d^4
l}{(2\pi)^4}\frac{\mathrm{Tr}[(l\!\!\!\slash+m_{s})\epsilon\!\!\!\slash_{2}\frac{(1+v\!\!\!\slash)}{2}
\epsilon\!\!\!\slash_{\gamma}\frac{(1+v\!\!\!\slash)}{2}\gamma_{5}\epsilon\!\!\!\slash_{1}]}{(l^2-m_{s}^2)(v\cdot
l+\Delta_{H})(v\cdot l+\Delta_{S})}\bigg\}\nonumber\\&&=
\sqrt{Z_{S}M_{2}Z_{H}M_{D_{s}^{*}}}\bigg(\frac{2e}{3}\bigg)
\varepsilon_{\alpha\beta\rho\sigma}q^{\alpha}\epsilon_{\gamma}^{\beta}
\epsilon_{1}^{\rho}\epsilon_{2}^{\sigma}\bigg[\mathcal{R}m_{s}-\mathcal{R}_{1}+\frac{2\mathcal{R}_{2}(p'\cdot
q)}{M_{D_{s}^{*}}}+
\frac{\mathcal{R}m_{s}^{2}-\mathcal{R}_{3}-2\mathcal{R}_{6}}{M_{D_{s}^{*}}}
\nonumber\\&&+ \frac{(2\mathcal{R}_{5}+2\mathcal{R}_{1})(p'\cdot
q)}{M^{2}_{D_{s}^{*}}}\bigg].
\end{eqnarray}

\vspace{0.5cm}

(d) $D_{sJ}(2460)\rightarrow D_{sJ}^{*}(2317)+\gamma$.

The transition matrix element of $D_{sJ}(2460)\rightarrow
D_{sJ}^{*}(2317)+\gamma$ is

\begin{eqnarray}
&&\mathcal{M}[D_{sJ}(2460)\rightarrow D_{sJ}^*(2317)+\gamma
]\nonumber\\&&= (-1)i^6
\sqrt{Z_{S}M_{1}Z_{S}M_{{2}}}N_{c}\bigg\{\bigg(\frac{-e}{3}\bigg)\int^{reg}\frac{d^4
l}{(2\pi)^4}\frac{\mathrm{Tr}[(l\!\!\!\slash+m_{s})\epsilon\!\!\!\slash_{\gamma}
(l\!\!\!\slash+q\!\!\!\slash+m_{s})
\frac{(1+v\!\!\!\slash)}{2}\gamma_{5}\epsilon\!\!\!\slash_{1}]}{(l^2-m_{s}^2)[(l+q)^2-m_{s}^2](v\cdot
l+\Delta_{S})}\nonumber\\&&+\frac{2e}{6}\int^{reg}\frac{d^4
l}{(2\pi)^4}\frac{\mathrm{Tr}[(l\!\!\!\slash+m_{s})\frac{(1+v\!\!\!\slash)}{2}
\epsilon\!\!\!\slash_{\gamma}\frac{(1+v\!\!\!\slash)}{2}\gamma_{5}\epsilon\!\!\!\slash_{1}]}{(l^2-m_{s}^2)(v\cdot
l+\Delta_{S})(v\cdot l+\Delta_{S})}\bigg\}\nonumber\\&&
=\sqrt{Z_{S}M_{1}Z_{S}M_{{2}}}\frac{2
e}{3}\frac{m_{s}\mathcal{R}}{M_{1}}\varepsilon_{\alpha\beta\rho\sigma}q^{\alpha}
p'^{\beta}\epsilon_{\gamma}^{\rho}\epsilon_{1}^{\sigma}.
\end{eqnarray}

The concrete expressions of $\mathcal{R}$ and $\mathcal{R}_{i}$
are listed in the appendix.

 \vspace{0.5cm}

\section{Numerical results}

With the formulation we derived in last section, we can
numerically evaluate the corresponding decay rates. Besides, we
need several input parameters for the numerical computations. They
include: $f_{\pi}=132$ MeV, $m_s=0.5$ GeV, $\Lambda=1.25$ GeV, the
infrared cutoff $\mu=0.51$ GeV and $\Delta_{S}-\Delta_{H}=335\pm
35$ MeV \cite{parameter}. $m_{\eta}=547.45$ MeV, $M_{1}=2317$ MeV,
$M_{2}=2460$ MeV, $M_{D_{s}}=1968$ MeV and $M_{D_{s}^*}=2112$ MeV
\cite{PDG}. The suppression parameter was estimated in
ref.\cite{coe} as
$$\frac{\tilde{m}_u -\tilde{m}_d
}{\tilde{m}_{s}-(\tilde{m}_{u}+\tilde{m}_{d})/2}\sim
\frac{1}{43.7}.$$

(a) We present the decay widths of $D_{sJ}^{*}(2317)\rightarrow
D_{s}+\pi^0$ and $D_{sJ}(2460)\rightarrow D_{s}^{*}+\pi^0$ in
Table \ref{strong}.
\begin{table}[htb]
\begin{center}
\begin{tabular}{cccccc} \hline
$\Delta_{H}(\mathrm{GeV})$&$\Delta_{S}(\mathrm{GeV})$ & $Z_{H}$
$(\mathrm{GeV})^{-1}$
 & $Z_{S}$ $(\mathrm{GeV})^{-1}$& $\Gamma_{1}$(keV)&$\Gamma_{2}$(keV)\\\hline
0.5&0.86&3.99&2.02&3.68&1.86\\\hline
0.6&0.91&2.69&1.47&5.36&2.72\\\hline
0.7&0.97&1.74&0.98&8.71&4.42\\\hline
\end{tabular}
\end{center}
\caption{The values of $\Delta_{S}$ and $\Delta_{H}$ are taken
from Ref. \cite{parameter}. According eqs. (\ref{ZH}) and
(\ref{ZS}), one gets the values of $Z_{S}$ and $Z_{H}$.
$\Gamma_{1}$ and $\Gamma_{2}$ denote respectively the decay widths
of $D_{sJ}^{*}(2317)\rightarrow D_{s}+\pi^0$ and
$D_{sJ}(2460)\rightarrow D_{s}^{*}+\pi^0$.}\label{strong}
\end{table}

For a comparison, we also list the values of the decay widths of
$D_{sJ}^{*}(2317)\rightarrow D_{s}+\pi^0$ and
$D_{sJ}(2460)\rightarrow D_{s}^{*}+\pi^0$ which are calculated by
other groups, in Table \ref{different}.
\begin{table}[htb]
\begin{center}
\begin{tabular}{c|c c c c c c c  } \hline
& CQM model
&\cite{Nielsen}&\cite{zhu}&\cite{decay-1}&\cite{decay-2}&\cite{decay-3}
&\cite{decay-4}\\\hline $ \Gamma_{1}$(keV)& $3.68\sim
8.71$&$6\pm2$ &$34\sim44$&$7\pm1$&21.5&$\sim 10$&$16$
\\\hline$
\Gamma_{2}$(keV)&$1.86\sim4.42$&-&$35\sim51$&$7\pm1$&21.5&$\sim10$&$32$
\\\hline
\end{tabular}
\end{center}
\caption{In this table, we list our calculation of
$D_{sJ}^{*}(2317)\rightarrow D_{s}+\pi^0$ and
$D_{sJ}(2460)\rightarrow D_{s}^{*}+\pi^0$ and that obtained by
other groups. Where $ \Gamma_{1}$ and $ \Gamma_{2}$ are
respectively denote the decay widths of
$D_{sJ}^{*}(2317)\rightarrow D_{s}+\pi^0$ and
$D_{sJ}(2460)\rightarrow D_{s}^{*}+\pi^0$.}\label{different}
\end{table}

\vspace{0.2cm}

Our values depend on the model parameters, but qualitatively, the
order of magnitude is unchanged when the parameters vary within a
reasonable region. All the values in table \ref{different} are
somehow consistent with each other for order of magnitude, even
though there is obvious difference in numbers.\\

(b) With the same parameters, we obtain the radiative decay rates
of $D_{sJ}^{*}(2317)$ and $D_{sJ}(2460)$. The results are listed
in Table. \ref{radiative}
\begin{table}[htb]
\begin{center}
\begin{tabular}{c|ccccc} \hline
& CQM model &
\cite{radi-LCQSR}&\cite{decay-1}&\cite{decay-2}&\cite{decay-3}\\\hline
 $\Gamma(D_{sJ}^{*}(2317)\rightarrow D_{s}+\gamma)$(keV)&$\sim 1.1$ &$4\sim 6$&$0.85$&$1.74$&$1.9$
\\\hline $\Gamma(D_{sJ}(2460)\rightarrow D_{s}+\gamma)$(keV)&$0.6\sim 2.9$ &$19\sim
29$&$3.3$&$5.08$&$6.2$
\\\hline $\Gamma(D_{sJ}(2460)\rightarrow D_{s}^{*}+\gamma)$(keV)&$0.54\sim 1.4$
&$0.6\sim1.1$&$1.5$&$4.66$&$5.5$
\\\hline $\Gamma(D_{sJ}(2460)\rightarrow D_{sJ}^{*}(2317)+\gamma)$(keV)&$0.13\sim0.22$ &
$0.5\sim 0.8$&-&$2.74$&$0.012$
\\\hline
\end{tabular}
\end{center}
\caption{The rates of $D_{sJ}^{*}(2317)\rightarrow D_{s}+\gamma$,
$D_{sJ}(2460)\rightarrow D_{s}+\gamma$, $D_{sJ}(2460)\rightarrow
D_{s}^{*}+\gamma$ and $D_{sJ}(2460)\rightarrow
D_{sJ}^{*}(2317)+\gamma$. Meanwhile we also list the results
obtained by other group.}\label{radiative}
\end{table}

\begin{table}[htb]
\begin{center}
\begin{tabular}{c|cccc} \hline
&Belle&Babar&CLEO \cite{CLEO}&CQM model \\ \hline
$R_{1}$&$< 0.18$\cite{belle}&-&$<0.059$&$0.12\sim 0.30$\\
\hline $R_{2}$&$0.55\pm 0.13\pm 0.08$ \cite{belle}&$0.375\pm
0.054\pm0.057$ \cite{babar-1}&$<0.49$&$0.32\sim 0.66$\\\hline
$R_{3}$&$<0.31$ \cite{belle}&-&$<0.16$& $0.29\sim0.32$
\\\hline $R_{4}$&-&$<0.23$ \cite{babar-1} &$<0.58$&
$0.05\sim0.07$
\\\hline
\end{tabular}
\end{center}
\caption{The ratios of radiative decay widths to strong decay
widths for $D_{sJ}^{*}(2317)$ and $D_{sJ}(2460)$. The first three
columns are the experimental data and the fourth column is our
result calculated in the CQM model.}
\end{table}

For convenience, we define the relevant ratios as
$$R_1=\Gamma(D_{sJ}^{*}(2317)\rightarrow
D_{s}+\gamma):\Gamma(D_{sJ}^{*}(2317)\rightarrow D_{s}+\pi^{0}),$$
$$R_2=\Gamma(D_{sJ}(2460)\rightarrow
D_{s}+\gamma):\Gamma(D_{sJ}(2460)\rightarrow D_{s}^*+\pi^{0}),$$
$$R_3=\Gamma(D_{sJ}(2460)\rightarrow
D_{s}^{*}+\gamma):\Gamma(D_{sJ}(2460)\rightarrow
D_{s}^*+\pi^{0}),$$
$$R_4=\Gamma(D_{sJ}(2460)\rightarrow
D_{sJ}^{*}(2317)+\gamma):\Gamma(D_{sJ}(2460)\rightarrow
D_{s}^*+\pi^{0}).$$

Then we calculate the $R_i$'s and tabulate the results below.

\vspace{0.2cm}

\vspace{0.5cm}

\section{Conclusion and Discussion}

In this work, based on the assumption that $D_{sJ}^{*}(2317)$ and
$D_{sJ}(2460)$ are chiral parters of $D_{s}$ and $D_{s}^*$, we
calculate the  rates of $D_{sJ}^{*}(2317)\rightarrow D_{s}\pi^{0}$
and $D_{sJ}(2460)\rightarrow D_{s}^{*}\pi^0$ in the
Constituent-Quark-Meson (CQM) model and  take into account the
$\eta-\pi^0$ mixing mechanism  \cite{wise}. We also estimate the
rates of $D_{sJ}^{*}(2317)\rightarrow D_{s}^{*}+\gamma$,
$D_{sJ}(2460)\rightarrow D_{s}+\gamma$, $D_{sJ}(2460)\rightarrow
D_{s}^* +\gamma$ and $D_{sJ}(2460)\rightarrow
D_{sJ}^{*}(2317)+\gamma$ in the same model.

Comparing our results about the strong decay rates with that
obtained by other groups, we find that our results are reasonably
consistent with the values listed in table \ref{different}.

For the radiative decay, our results generally coincide with that
obtained by other groups and especially  these results of the QCD
sum rules.

The ratio $R_2$ has been measured with relatively high precision
\cite{exp1,belle,babar-1}, however for  $R_1,\; R_3$ and $R_4$,
there only are upper limits given by Babar, Belle, and CLEO
collaborations\cite{CLEO,belle,babar-1}. It seems that our results
on the ratios well coincide with the experimental values. This
consistency somehow implies that  the assumption of
$D_{sJ}^*(2317), \;D_{sJ}(2460$ being p-wave chiral partners of
$D_s,\; D_s^*$ does not contradict to the data with the present
experimental accuracy.

The experimental upper bounds on the total widths are
$\Gamma(D_{sJ}^*(2317))< 4.6$ MeV and $\Gamma(D_{sJ}(2460))< 5.5$
MeV \cite{PDG}. Obviously, the overwhelming decay modes of
$D_{sJ}^*(2317)$ and $D_{sJ}(2460)$ are their strong and radiative
decays, therefore we can roughly take sums of these widths as the
total widths of $D_{sJ}^*(2317)$ and $D_{sJ}(2460)$. However, our
numerical results as well as that given by other groups are in
order of tens of keV,  much smaller than the upper bounds set by
recent experiments. The reason is obvious that the aforementioned
reactions violate isospin conservation, there is a large
suppression factor of about $(1/43.7)^2$, which reduces the widths
by 3 orders. The calculations which are based on the assumption
that the newly discovered $D_{sJ}^*(2317)$ and $D_{sJ}(2460)$ are
p-wave excited states of $D_s$ and $D_s^*$ predict their widths to
be at order of a few to tens of keV. By contrary, if they are
four-quark states, or molecular states, there may be more decay
channels available, i.e. some modes are not constrained by the OZI
rule, thus much larger total widths might be expected in that
scenario. The authors of refs.\cite{decay-5,decay-6}, for example,
suggested that $D_{sJ}^{*}(2317)$ and $D_{sJ}(2460)$ are in
non-$c\bar s$ structure (four-quark states etc.), and obtained
much larger rates, even though still obviously smaller than the
experimental upper bounds. So far, it is hard to conclude if they
are p-wave chiral partners of $D_s$ and $D_s^*$ or four-quark
states yet.

We  hope that the further more precise measurements of Babar,
Belle and CLEO may offer more information  by which we may
determine the structure of the newly discovered mesons.

\vspace{0.5cm}

\noindent Acknowledgment:

We would like to thank Dr. Xin-Heng Guo for helpful discussions.
This work is supported by the National Natural Science Foundation
of China (NNSFC).

\vspace{1.5cm}

 {\bf{Appendix}}

\begin{eqnarray}
\mathcal{O}&=&\frac{I_{5}(\Delta_{S},m_{\eta}/2,
\omega)-I_{5}(\Delta_{H},-m_{\eta}/2, \omega)}{2m_{\eta}},\\
\mathcal{O}_{1}&=&\frac{I_{3}(-m_{\eta}/2)-I_{3}(m_{\eta}/2)+\omega
[I_{3}(\Delta_{S})-I_{3}(\Delta_{H})]}{2m_{\eta}(1-\omega^2)}-\frac{\mathcal{O}
[\Delta_{S}-\omega m_{\eta}/2]}{1-\omega^2},\\
\mathcal{O}_{2}&=&\frac{-I_{3}(\Delta_{S})+I_{3}(\Delta_{H})-\omega
[I_{3}(-m_{\eta}/2)-I_{3}(m_{\eta}/2)]}{2m_{\eta}(1-\omega^2)}-\frac{\mathcal{O}
[m_{\eta}/2-\Delta_{S}\omega]}{1-\omega^2},\\
\mathcal{O}_{3}&=&\frac{\mathcal{B}_{1}}{2}+\frac{2\omega
\mathcal{B}_{4}-\mathcal{B}_{2}-\mathcal{B}_{3}}{2(1-\omega^2)^2},\\
\mathcal{O}_{4}&=&\frac{-\mathcal{B}_{1}}{2(1-\omega^2)}+\frac{3\mathcal{B}_{2}-6\omega
\mathcal{B}_{4}+\mathcal{B}_{3}(2\omega^2+1)}{2(1-\omega^2)^2},\\
\mathcal{O}_{5}&=&\frac{-\mathcal{B}_{1}}{2(1-\omega^2)}+\frac{3\mathcal{B}_{3}-6\omega
\mathcal{B}_{4}+\mathcal{B}_{2}(2\omega^2+1)}{2(1-\omega^2)^2},\\
\mathcal{O}_{6}&=&\frac{\mathcal{B}_{1}\omega}{2(1-\omega^2)}+\frac{2\mathcal{B}_{4}
(2\omega^2+1)-3\omega(\mathcal{B}_{2}+\mathcal{B}_{3})}{2(1-\omega^2)^2},\\
\mathcal{B}_{1}&=&m_{s}^2 \mathcal{O}-I_{3}(\Delta_{H}),\\
\mathcal{B}_{2}&=&\Delta^2_{S}\mathcal{O}-\frac{I_{3}(m_{\eta}/2)-I_{3}(-m_{\eta}/2)}
{4m_{\eta}}(\omega m_{\eta}+2\Delta_{S}),\\
\mathcal{B}_{3}&=&\frac{m_{\eta}^2\mathcal{O}}{4}+\frac{I_{3}(\Delta_{S})
-3I_{3}(\Delta_{H})}{4}+\frac{\omega}{4}[\Delta_{S}I_{3}(\Delta_{S})-\Delta_{H}I_{3}
(\Delta_{H})],\\
\mathcal{B}_{4}&=&\frac{m_{\eta}\Delta_{S}\mathcal{O}}{2}+\frac{\Delta_{S}
[I_{3}(\Delta_{S})
-I_{3}(\Delta_{H})}{2m_{\eta}}+\frac{I_{3}(m_{\eta}/2)-I_{3}(-m_{\eta}/2)}{4},\\
I_{5}(a_{2},a_{3},a_{4})&=&\int^{1}_{0}d
x\frac{1}{1+2x^2(1-a_{4})+2x(a_{4}-1)}\bigg\{\frac{6}{16\pi^{3/2}}
\int^{1/\mu^2}_{1/\Lambda^2}\frac{d y}{\sqrt{y}} \xi\exp[-y(m_{s}^{2}-\xi^2 )]\nonumber\\
&&\times[1+\mathrm{erf}(\xi \sqrt{y})]+\frac{6}{16\pi^2}
\int^{1/\mu^2}_{1/\Lambda^2}\frac{d y}{{y}}\exp[-y(m_{s}^{2}-2\xi^2 )]\bigg\},\\
\xi=&=&\frac{a_{2}(1-x)+a_{3}x}{\sqrt{1+2(a_{4}-1)x+2(1-a_{4})x^2}}.
\end{eqnarray}

\begin{eqnarray}
\mathcal{R}&=&\frac{3}{16\pi^{3/2}}\int^{1/\mu^2}_{1/\Lambda^2}ds\frac{\exp[-sm_{s}^2]}{s^{1/2}}\int^{1}_{0}
d x \exp[s\Delta^2(x)][1+\mathrm{erf}(\Delta(x)\sqrt{s})],\\
\mathcal{R}_{1}&=&\frac{n_{2}}{|\mathbf{q}|},\\
\mathcal{R}_{2}&=&\frac{n_{1}-\mathcal{R}_{1}}{|\mathbf{q}|},\\
\mathcal{R}_{3}&=&\frac{n_4}{|\mathbf{q}|^2},\\
\mathcal{R}_{6}&=&\frac{\mathcal{R}_3
+n_{6}}{2}-\frac{n_{5}}{|\mathbf{q}|},\\
\mathcal{R}_{5}&=&\frac{n_{5}}{|\mathbf{q}|^2}-\frac{\mathcal{R}_{3}+\mathcal{R}_{6}}{|\mathbf{q}|},\\
\mathcal{R}_{4}&=&\frac{n_{3}-\mathcal{R}_{3}-\mathcal{R}_{6}}{|\mathbf{q}|^2}-\frac{2\mathcal{R}_{5}}{|\mathbf{q}|},\\
\Delta(x)&=&\Delta_{S}-x|\mathbf{q}|,\;\;\;I_{2}=\frac{3}{16\pi^2}\Gamma(0,\frac{m_{s}^2}{\Lambda^2},\frac{m_{s}^2}{\mu^2})\\
n_{1}&=&-I_{2}-\Delta_{S}\mathcal{R},\\
n_{2}&=&\frac{1}{2}[-I_{3}(\Delta_{S})+I_{3}(\Delta_{H})],\\
n_{3}&=&\Delta_{S}^{2}\mathcal{R}+\Big(\frac{|\mathbf{q}|}{2}+\Delta_{S}\Big)I_{2},\\
n_{4}&=&\frac{|\mathbf{q}|}{2}[\Delta_{S}I_{3}(\Delta_{S})-\Delta_{H}I_{3}(\Delta_{H})],\\
n_{5}&=&\frac{\Delta_{S}}{2}[I_{3}(\Delta_{S})-I_{3}(\Delta_{H})],\\
n_{6}&=&m_{s}^2\mathcal{R}-I_{3}(\Delta_{H}),
\end{eqnarray}
where $\mathbf{q}$ is the three momentum of the emitted photon.

\end{document}